\documentstyle[prince,epsf,11pt]{article}
\def\captionfontsize{\small}
\def\tablefontsize{\footnotesize}
\def\meanl{\mbox{$<$}}
\def\meanr{\mbox{$>$}}
\def\kms{\nobreak\mbox{$\;$km\,s$^{-1}$}}

\def\la{\mathrel{\hbox{\rlap{\hbox{\lower4pt\hbox{$\sim$}}}\hbox{$<$}}}}
\def\ga{\mathrel{\hbox{\rlap{\hbox{\lower4pt\hbox{$\sim$}}}\hbox{$>$}}}}
\begin{document}
\thispagestyle{empty}
\title{THE EVIDENCE FOR THE LONG DISTANCE SCALE WITH H$_{0} <$ 65}
\author{Allan Sandage\\[0.5ex]
 The Observatories of the Carnegie Institution of Washington\\[1.5ex]
                           and\\[1.5ex]
                       G.A.Tammann\\[0.5ex]
         Astronomisches Institut der Universitat Basel}

\begin{abstract}\small
The status of the determination of the Hubble constant
is reviewed, setting out the evidence for the long distance scale 
with $H_{0}= 55 \pm 5$. In parallel, various precepts used by others, 
said to favor the short distance scale with $H_{0} > 70$ are discussed. 
The strongest evidence for the long scale are (1) the calibration of 
the peak absolute magnitude of type Ia supernovae with their 
Hubble diagram tied to the remote cosmic kinematic frame, (2) the 
distance to the Virgo cluster by six largely independent methods 
including Tully-Fisher using a complete cluster sample and a new 
calibration using recent HST Cepheid data, and (3) field spirals 
binned by luminosity class, also calibrated using Cepheid 
distances. The three methods give 
$H_{0} = 56 \pm 3, 55 \pm 2$, and $53 \pm 3$ 
(internal errors). $H_{0}$ does not vary significantly over scales 
from 10-500 Mpc. $H_{0}$ does not increase outward, as appearances 
using field galaxies would give if the raw data were not 
corrected for observational selection bias.      

     Higher values of $H_{0}$ still in the literature are based on (1)
an untenably small distance to the Virgo cluster claimed by 
equating (against newly available evidence) the Cepheid distance 
of M100 with the mean distance of the cluster, (2) an untenably 
large Virgo cluster velocity tied to the remote cosmic kinematic 
frame, (3) a questionable route through the Coma cluster on the 
assumption that its random motion can be neglected at its 
assumed distance, (4) an incorrect precept that the Cepheid 
distance to NGC 1365, a possible member of the Fornax cluster, 
gives the distance to NGC\,1613, parent to two SNe\,Ia, calibrating 
them, (5) an unjustified reliance on planetary nebulae and 
surface brightness fluctuations as distance indicators at the 
present stage of their calibration, and (6) either an 
underestimation or a neglect of the importance of observational 
selection bias in flux-limited samples, both for cluster galaxies 
(the Teerikorpi cluster incompleteness bias), or for field 
galaxies (the Malmquist bias). There is no valid evidence for       
$H_{0} > 70$.  

     The status of the time scale test is reviewed using recent
discussions of the age of globular clusters based on seven 
studies since 1993. The result is 13-14 ( $\pm 2$) Gyr for the age of 
the Galactic globular cluster system. Even with a gestation 
period of the Galaxy of 1 Gyr, there is no time scale crisis in 
cosmology provided that 
$q_{0} < 0.3$, $H_{0} = 55$, and $\Lambda  = 0$.     
\end{abstract}

\section{The Controversy}\label{sec:1}

     It is written that practical cosmology reduces to the 
``search for two numbers". This simplicity was pronounced before 
the marriage had occurred between high-energy particle physics 
in the free quark era and the actual cosmological observations 
made at the telescope. Nevertheless, one of the few premises 
still agreed to by the current debaters is that the value of the 
Hubble constant remains central to the subject. 

     The most important reason now, even as in Hubble's time,
concerns the time scale. If the inverse Hubble constant were, for 
real, smaller than a ``known age of the universe" (significantly 
outside the errors), then the standard model, sans cosmological 
constant, falls.      

     Since 1978 our critics have espoused Hubble constants that
began with values larger than 100\kms\ Mpc$^{-1}$ (de Vaucouleures 
\& Bollinger 1979 with earlier references therein), found by using 
(1) what are now known to be incorrect local distance 
calibrations, (2) by not correcting for selection bias, or (3) by 
claiming the absence of bias altogether and a Hubble constant 
that increases outward (de Vaucouleurs \& Peters 1986). This is a 
sure signature that bias in flux-limited samples exists (Sandage 
1988a, 1994a). 

     Although the values of $H_{0}$ by the proponents of the short
distance scale have gradually decreased over time, yet most 
claims even by commentators that themselves have not been in the 
arena (Fukugita et al.~1993; Hogan 1994; Bolte \& Hogan 1995) 
are for $H_{0}$ between 70 and 85, giving $H_{0}^{-1}$ between 
14 and 11.5 Gyr. Recall that the free-expansion age is 
$H_{0}^{-1}$ = 9.8 Gyr for $H_{0}$ = 100. 
These numbers would, of course, support a time-scale 
crisis even for an empty Universe with objects of ages $> 14$ Gyr.

     We began a series of investigations on the distance scale in 
1963 with the Palomar Hale telescope, first measuring the Cepheid 
distance to NGC 2403 (Tammann \& Sandage 1968), and continuing in 
a series of papers called ``Steps Toward the Hubble Constant" 
(Sandage \& Tammann 1974 for Paper I; 1995a for Paper X). We are 
now proceeding again (cf. Saha et al.~1997 for Paper VIII of a 
new series, with earlier references therein) using Cepheid 
distances to type Ia supernovae measured with HST. Our value of 
the Hubble constant has consistently been steady near $H_{0} = 55   
\pm  10$ since 1974.      

     Nevertheless, our low value of $H_{0}$ has been generally
discounted. The principal reason is the quite astounding 
apparent internal agreement of the several new and 
independent methods amongst themselves (Tully-Fisher, planetary 
nebulae, surface brightness fluctuations, globular clusters) used 
by others beginning in the mid 1980's, generally giving $H_{0}$ 
between 80 and 100. Part of the agreement is, of course, a 
lemming effect, where, when a choice between precepts giving 
different final values must be made, that choice has often gone 
to the high $H_{0}$ value because other methods appeared to be in 
support. But that support could be shown either to be soft, or in 
fact, incorrect.    

     To that point, we have often reviewed the subject, both in
conference reports or monographs (Tammann 1986, 1987, 1988, 1992, 
1993, 1996a,b; Tammann \& Sandage 1996; Sandage \& Tammann, 1982, 
1984, 1986, 1995b; Sandage 1993, 1995b, 1996) and in 
Journal papers (Tammann \& Sandage 1995; Sandage \& Tammann 1995a, 
1996; Sandage 1988a,b; 1994a,b; 1996a,b; Federspiel, Sandage, \& 
Tammann 1994; Sandage, Tammann, \& Federspiel 1995), setting out 
the reasons why each of the methods said to give the high values 
of $H_{0}$ contain the same types of error, generally traced either to 
(1) neglect of the pernicious effect of observational selection 
bias, (2) incorrect panegyrics of why particular samples and/or 
methods are immune from, and therefore need not be corrected for, 
bias, or (3) a misunderstanding of methods to correct for the 
bias even when the need for correction is clear. 

     In each case, we have shown that the application of bias
corrections, and/or a more proper calibration of the methods 
themselves (in particular Tully-Fisher and globular clusters), 
reduce the high values of $H_{0}$ to less than 65. 

     The same conclusions, for nearly the same reasons, have also
been reached by Bottinelli et al.~(1986a,b; 1987), Teerikorpi 
(1987), and now by Theureau et al.~(1996) in which they obtain      
$H_{0}$ = 55 using their large sample of field galaxies with the 
Tully-Fisher method, calibrated with the recent Cepheid distances, 
carefully corrected for selection bias.  

     The purpose of this report is to update the current state of
the debate. The plan of the paper is to (1) set out in 
Section~\ref{sec:2} the fact that the Hubble constant exists and 
that its rate can be determined beyond all local perturbations of 
the velocity field by tying $H_{0}$ to the remote cosmic kinematic 
frame, (2) to present in Section~\ref{sec:3} and Section~\ref{sec:4} 
the evidence based on methods using type Ia supernovae, 21 cm line
widths, and globular clusters. All three give $m - M = 31.7$ 
for the distance modulus of the Virgo cluster. This distance 
($D = 22\;$Mpc), when tied into the proper Virgo cluster cosmic 
velocity frame, gives $H_{0} = 54 \pm  5$. (3) In Section~\ref{sec:5} 
we review the method used by Freedman et al.~(1994), 
Tanvir et al.~(1994), and Whitmore et al.~(1995) in their adoption 
of a demonstrably incorrect distance modulus to the Virgo cluster 
core, and that by stepping this distance to Coma, obtain an 
incorrectly high value of $H_{0}$. (4) We discuss in  
Section~\ref{sec:6} the recent route through the Fornax cluster 
(Freedman et al.~1996) that assumes that the modulus of NGC\,1365 
defines the distance to NGC\,1316, parent galaxy to two normal 
SNe\,Ia, thereby incorrectly calibrating $\meanl M(\max)\meanr$ 
for SNe\,Ia. (5) We set out in Section~\ref{sec:8} the 
independent evidence using field galaxies, corrected for 
selection effects and calibrated using Cepheids in local 
galaxies, that $H_{0} = 53 \pm 3$. (6) In Section~\ref{sec:9} 
we comment on physical methods that are independent of the distance 
scale ladder, also leading to $H_{0} \approx 50\!-\!65$, and (7) 
finally set out in Section~\ref{sec:10} the current 
position on the time scale test of the standard model.

\section{The Hubble constant exists} \label{sec:2}

     The Hubble constant means something only if the expansion 
is real and if the form of the expansion velocity-field is 
linear.\footnote{%
   The term ``velocity of expansion" appears to be meaningless 
   in cosmology, the expansion being a time variation of the metric 
   scale factor in the famous Lemaitre equation, not a ``Doppler" 
   effect (Harrison 1981). Of course, that fraction of the redshift 
   that reflects peculiar (random and streaming) motions does, 
   undoubtedly, mean real Doppler velocities.}   
There is now no question that both of these conditions
are met. 

     Consider first the reality of the expansion. Three tests exist.
Each has proved positive. (a) The Tolman surface brightness (SB)   
test that identical luminous objects will have SB's that become 
fainter with redshift as $(1 + z)^{4}$, appears to have been verified 
(Sandage \& Perelmutter 1991; see also Kjaergaard et al.~1993 for 
a modified method). (b) The time dilution test, based on the 
prediction of Wilson (1939) that standard clocks will appear to 
be slow by the factor (1 + z), has also apparently been verified 
(Perlmutter et al.~1995) using SNe\,Ia, (3) the temperature of the 
relic radiation must increase as (1 + z) (Tolman 1934, eq. 171.6 
combined with 171.2). The effect has apparently been measured by 
Songaila et al.~(1994), replacing the upper limits known before 
(Meyer et al.~1986).    

     That the form of the Hubble expansion is linear with
distance has been proved by using progressively more suitable 
``standard" objects in the Hubble diagram such as brightest 
cluster galaxies (BCG), and/or SNe of type Ia. A linear redshift-
distance relation is proved by a straight line correlation in 
that diagram between apparent magnitude and log redshift with a 
slope of d\,mag$/$d\,$\log v = 5$ required by the inverse square law 
for intensity diminution with distance. The scatter will be small 
only if the spread in absolute luminosity is small. Clearly, in 
the absence of random motions, the scatter, read as residuals in 
apparent magnitude at a given redshift, measures the spread in 
absolute magnitude.  

     To determine the absolute value of distances in such Hubble
diagrams, giving the Hubble constant when the distances are 
divided into the observed redshifts, requires only that any 
particular Hubble diagram be calibrated using absolute 
magnitudes. In what follows we calibrate two such diagrams, one 
based on normal SNe\,Ia, and the other using distance {\em ratios} 
to Virgo, plus the absolute distance to Virgo determined by several 
methods.        

     The proof that the expansion is linear began with the first
extensive data by Hubble \& Humason (1931) that enlarged the 
original sample of Hubble (1929) many fold. It continued through 
the Palomar/Stromlo campaign of the 1960-1980's where many 
clusters were added around the sky (Sandage 1972, 1975, 1978; 
Sandage \& Hardy 1973) to test for isotropy of the 
expansion.\footnote{%
   The requirement that the form of the expansion must be linear 
   in the standard model, although perhaps obvious, is discussed in 
   some detail elsewhere (Sandage 1995a, Lecture 3).}  

\begin{figure}[h]
\begin{center}
\leavevmode
\epsfxsize 14.5cm
\epsffile{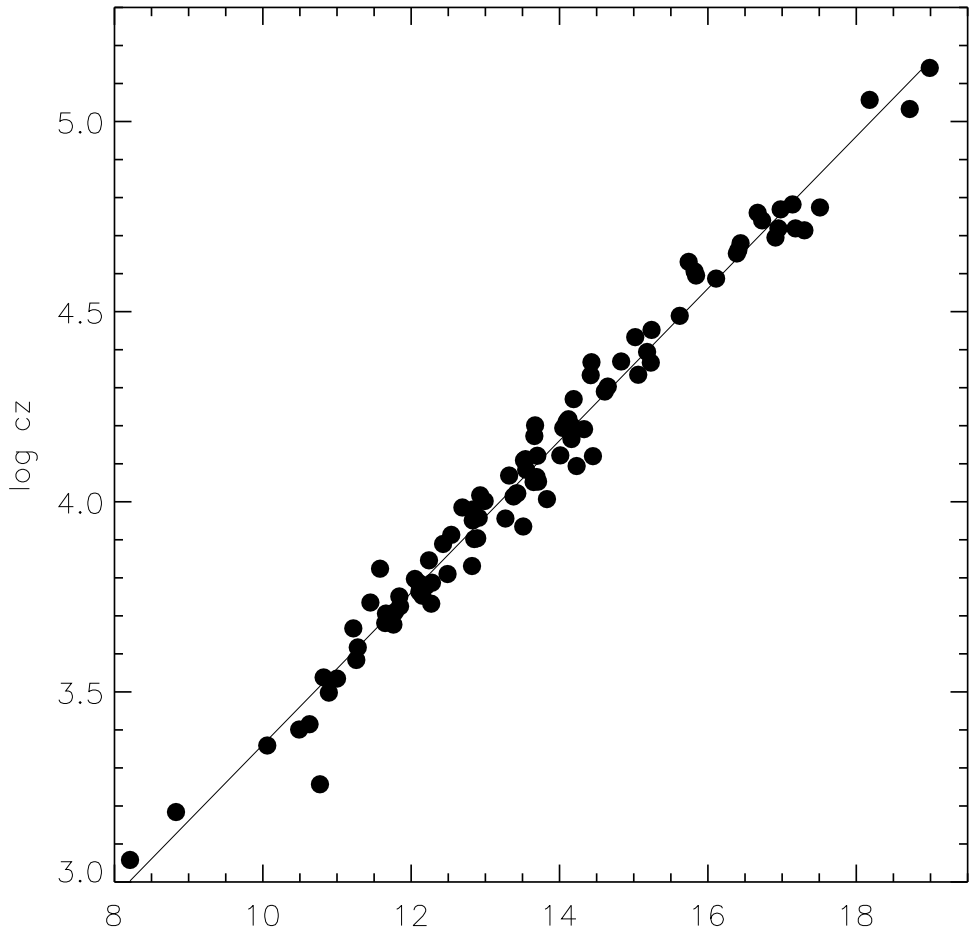}
%
\caption{The Hubble diagram using brightest galaxies in clusters 
     and groups with data determined at Palomar and Stromlo. The 
     abscissa is apparent $V$ magnitude corrected to a standard 
	 metric size, for K dimming, for Galactic absorption, for 
	 Bautz-Morgan contrast effect, and for the population richness 
	 effect. (Diagram from Sandage \& Hardy 1973).} \label{fig:1}
\end{center}
\end{figure}

     Figure~\ref{fig:1} shows the Hubble diagram made at an
%
%
intermediate stage in the Palomar program, using brightest
cluster galaxies (BCG) (see also Sandage, Kristian, \& Westphal 
1976). The line has the forced slope of d\,mag$/$d\,$\log\,z = 5$. 
Clearly this requirement for a linear expansion is satisfied directly 
from the data.    

\begin{figure}
\begin{center}
\leavevmode
\epsfxsize 13.0cm
\epsffile{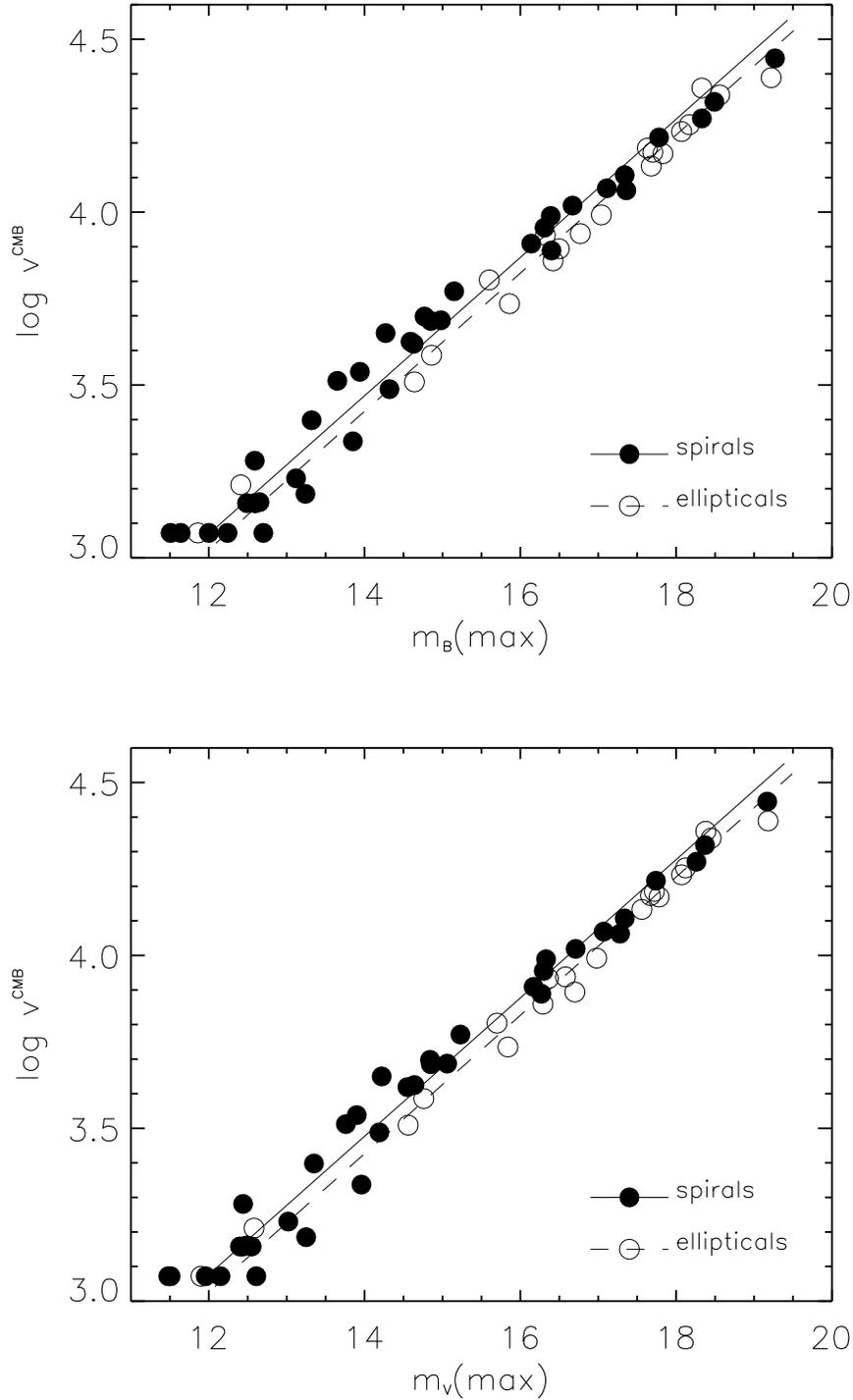}
\caption{Hubble diagrams in $B$ (top) and $V$ (bottom) for 56 blue 
     SNe\,Ia with $B(\max) - V(\max) \le 0.20$ after correction for 
	 Galactic absorption. The linear regressions for {\em all\/} 
	 SNe\,Ia give $m_{\rm B}(\max)=5\,\log v - (3.26 \pm 0.04)$ and 
	 $m_{\rm V}(\max)=5\,\log v - (3.30 \pm 0.04)$. (Data from Hamuy 
	 et al.~1996, Riess 1996, Patat 1996, Leib\-undgut et al.~1991
     and some additional sources).} \label{fig:2}
\end{center}
\end{figure}

     Fig.~\ref{fig:2} shows the Hubble diagram for a complete
%
%
sample of 56 supernovae of type Ia at 
maximum light in $B$ (top) and $V$ (bottom). Again the line is forced 
to have a slope of d\,mag$/$d\,log $v = 5$, and clearly the fit of the 
data to the line is excellent. The sample is selected from the 
literature to have reasonably well determined $B$(max) and $V$(max)
magnitudes, $3<\log v<4.5$, and $B(\max) - V(\max) \le 0.20$.
The color restriction is to exclude highly absorbed and intrinsically
red SNe\,Ia, like SN\,1992K, which are known to be heavily underluminous.
All SNe\,Ia fullfilling the above color restriction with known spectrum
are spectroscopically ``Branch-normal" 
(cf. Branch, Fisher, \& Nugent 1993), the only exception being 
SN\,1991T which is also significantly brighter than other SNe\,Ia in 
the Virgo cluster. Its exclusion could only decrease the true value of 
$H_{0}$ (cf. Section 3.2). As seen in Figure~\ref{fig:2} SNe\,Ia in 
spirals appear to be brighter by $0.23 \pm 0.08$ in $B$ and 
$0.26 \pm 0.08$ mag in $V$ then those in E/S0 galaxies. 

     The second important point from Figs.~\ref{fig:1} and \ref{fig:2} 
is the very small dispersion about the line, showing (1) that the 
expansion velocity field is extraordinarily quiet, seen by reading the 
residuals vertically (see Tammann \& Sandage 1985; Sandage \& 
Tammann 1995b concerning velocity anomalies on all scales), and 
(2) that the spread in absolute magnitude for brightest cluster 
galaxies and SNe\,Ia is small, seen by reading the residuals 
horizontally.  

\begin{figure}
\begin{center}
\leavevmode
\epsfxsize 12.4cm
\epsffile{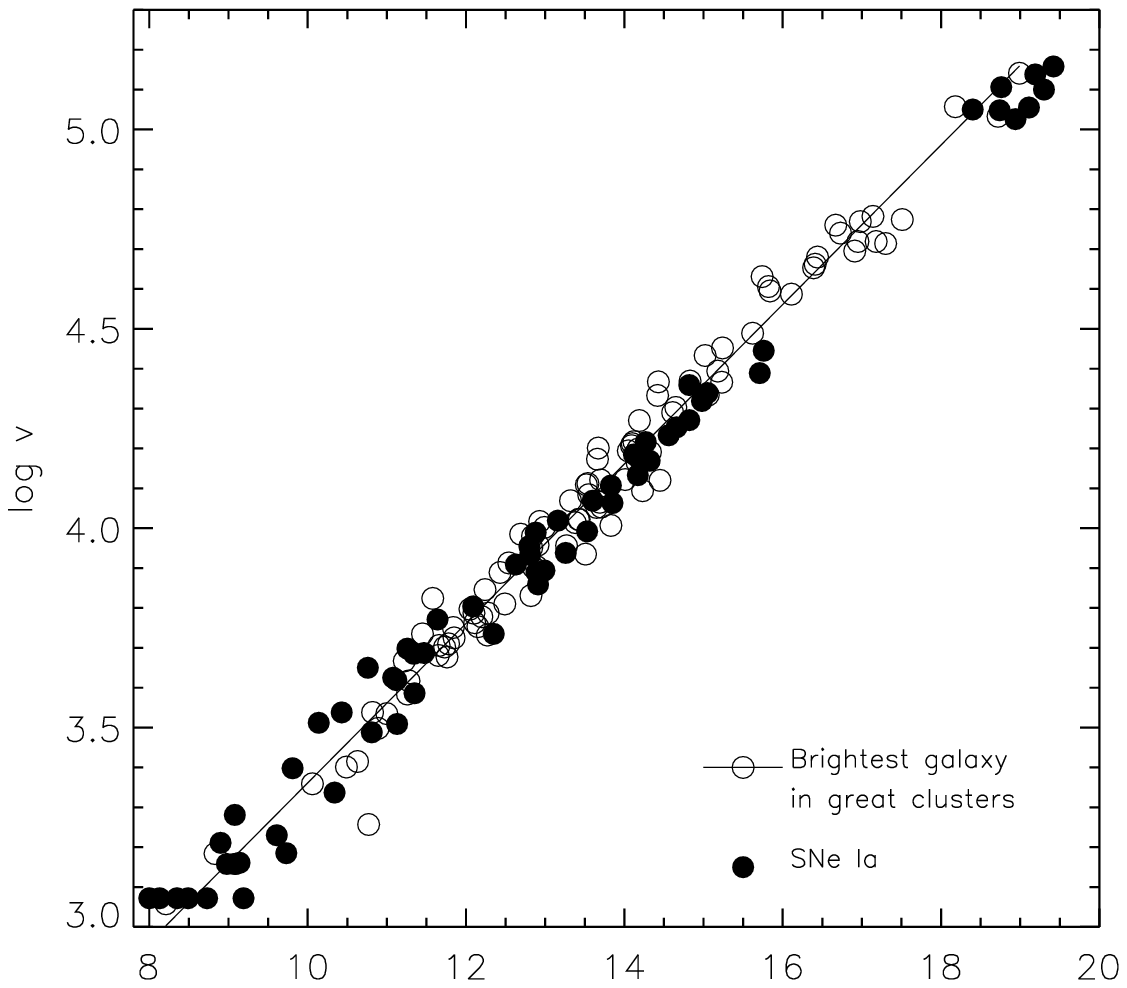}
\caption{Combination of Figs. 1 and 2 to emphasize 
     the smallness of the residuals about the line of forced slope 5, 
	 proving (1) the linearity of the expansion, (2) the lack of large 
	 streaming motions about the cosmic flow (vertical residuals), 
	 and (3) the sharply peaked luminosity function both for BCG and 
	 SNe\,Ia (horizontal residuals).
     The SNe\,Ia with $\log v>4.5$ were kindly provided by Perlmutter 
	 (1996) and Leibundgut et al.~1996). Note that $m_{\rm V}$  is 
	 plotted for BCGs but $m_{\rm B}$ for SNe\,Ia.}\label{fig:3}
\end{center}
\end{figure}

   Figure~\ref{fig:3} emphasizes the last point by combining the data in
Figures~\ref{fig:1} and \ref{fig:2} with an appropriate absolute 
magnitude offset. The small sigmas of the residuals, noted in the 
diagrams, show that BCG and SNe\,Ia are among the best standard 
candles known. 

     The Hubble diagram in Fig.~\ref{fig:3} can be read at large enough
redshifts (i.e. v $> 20000$\kms) to define the {\em global} value of 
$H_{0}$, freed from all velocity anomalies. $H_{0}$ follows from these 
diagrams once the absolute magnitude calibration of SNe\,Ia, or 
BCGs can be made. This has been accomplished in the manner set 
out in the next two sections. 

\section{Route through SNe\,Ia Calibrated by Cepheids} \label{sec:3}
 
     Our strongest evidence for the long distance scale is the 
determination of the absolute magnitude of Branch-normal SNe\,Ia, 
calibrating thereby Figure~\ref{fig:2}. The calibration is based on 
Cepheid distances to the parent galaxies. At this writing, 
Cepheids have been discovered, measured, and discussed in five 
nearby parent galaxies that have produced six SNe\,Ia. The ongoing 
experiment using Cepheids in galaxies that have produced SNe\,Ia 
is by a consortium composed of Sandage, Saha, Labhardt, 
Macchetto, Panagia and  Tammann.    

\subsection{Reliability of the Cepheid Zero Point} \label{sec:3_1}

     The Cepheids are noncontroversial as being the most reliable 
extragalactic distance indicators known. Agreement of the zero 
point of their P-L relation has been achieved to better than 0.1 
mag as determined over the past 30 years by a number of authors, 
including those shown in Table~\ref{tab:1}.  

%
\begin{table} 
\caption{The absolute $V$ magnitude of Cepheids with
         P = 10 days according to different P-L relations}\label{tab:1}
\begin{center} \tablefontsize
\begin{tabular}{lc}
               Source       &   $\meanl M_{\rm V}\meanr$ at 10 days  \\
\noalign{\smallskip}
\hline
\noalign{\smallskip}
             Kraft (1961)               &        -4.21  \\
             Sandage \& Tammann (1968)  &        -4.20  \\
             Feast \& Walker (1987)     &        -4.13  \\
             Madore \& Freedman (1991)  &        -4.14  \\
\end{tabular}
\end{center}
\end{table}
%

     The zero points of the first three entries have been 
determined by photometric parallaxes of Galactic clusters and 
associations containing Cepheids. As such, they ultimately rest 
on trigonometric parallaxes that define the zero point of the 
age-zero main sequence of the HR diagram. 

     When the zero point of the first three entries is applied to
Cepheids in the Large Magellanic Cloud, the distance modulus of 
LMC is {\em determined} to be $18.50 \pm 0.10$ 
(Sandage \& Tammann 1968; Feast \& Walker 1987). 
The zero point of Madore \& Freedman (1991), which 
we adopt in the following because these authors also give a 
calibration of the I band P-L relation, rests on the {\em assumption} 
that $(m - M)_{\rm LMC}^{0} = 18.5$. Is the assumption correct?  

     Independent confirmation of the Cepheid LMC modulus, and,
therefore, of the Cepheid zero point in Table~\ref{tab:1}, 
comes from five other methods, summarized elsewhere 
(Tammann 1996a, his Table 2).They give $(m - M)_{\rm LMC}^{0} = 18.57  
 \pm 0.06$, confirming the adopted zero point of the P-L relation 
within an error of less than 10\%. 

     The evidence is also strong that the zero point is virtually
independent of variations in metallicity over the range of [Fe/H] 
from 0 to -2 (Freedman \& Madore 1990; Chiosi et al.~1993; Sandage 
1996). Furthermore, selection bias, caused by the intrinsic 
spread of the P-L relation related to the intrinsic width of the 
instability strip, can also be avoided, but only if the Cepheids 
span a sufficient period range and if the data are suitably 
restricted in period (Sandage 1988c).  

\subsection{Calibration of SNe\,Ia Without Second-Parameter Effects} 
\label{sec:3_2}
     
     Rather than again setting out the detailed results of 
the current status of the SNe\,Ia experiments, summarized  
elsewhere (Saha 1996; Sandage et al.~1996; Tammann 1996a; 
Saha et al.~1997), we simply state the result. 

     The mean absolute magnitudes for Branch-normal SNe\,Ia, based
on either six (in $V$) or seven (in $B$) independent calibrations, 
are  
\begin{equation}\label{eq:1}
               M_{\rm B}(\max) = -19.52  \pm 0.07,          
\end{equation}
\begin{equation}\label{eq:2}
               M_{\rm V}(\max)  = -19.48  \pm 0.07.          
\end{equation}

     That these values conflict fundamentally with the short 
distance scale with $H_{0} \approx  85$ is seen by comparing 
equations~(\ref{eq:1}) and (\ref{eq:2}) with the statement by 
de Vaucouleurs (1979, his Table 9) that $\meanl M_{\rm V}(\max)\meanr$ 
must be $-$18.50 for his distance scale with 
$H_{0}$ = 88 to be correct, or the statement by Pierce (1994) that 
$\meanl M_{\rm B}(\max)\meanr$ must be $-$18.74 for his 
value of $H_{0}$ = 86 to be correct. 

     Equations~(\ref{eq:1}) and (\ref{eq:2}) show that the scale of 
de Vaucouleurs would be $H_{0}$ = 56 and that of Pierce would be 
$H_{0}$ = 61 if corrected to the values in equations~(\ref{eq:1}) 
and (\ref{eq:2}). Calculated in this way, the average would be 
\meanl $H_{0}\meanr = 58$. That value will, however, still 
contain the random errors of each of the methods used in the 
determinations by de~Vaucouleures and, independently, by Pierce 
to obtain their {\em relative\/} distances. 

     Combining equations~(\ref{eq:1}) and (\ref{eq:2}) with the 
equations of the ridge lines in Figure~\ref{fig:2} for blue 
SNe\,Ia gives Hubble constants of
\begin{equation}\label{eq:3}
               H_{0} = 56 \pm 2 \qquad\mbox{(internal error),}     
\end{equation}
and
\begin{equation}\label{eq:4}
               H_{0} = 58 \pm 2 \qquad\mbox{(internal error).}        
\end{equation}
 
      The results are quite robust against various subsamples 
of both the seven calibrating SNe\,Ia and the Hubble diagrams 
if they are divided into spiral and E galaxy groups and further 
separated by pre and post 1985 data.

      An interesting subsample are the 12 blue SNe\,Ia which have
occurred in spiral galaxies after 1985 in the distance range 
$3.8<\log v<4.5$. They match best the seven calibrating SNe\,Ia
which lie predominantly in spiral galaxies, they have the most
reliable photometry, and they are least affected by any peculiar
motions. They give with the calibration in equation~(\ref{eq:1})
and (\ref{eq:2}) $H_{0} = 55 \pm 2$ in $B$ and 
$H_{0} = 57 \pm 2$ in $V$.
Their rms scatter about the Hubble ridge line is only 
$\sigma_{B}=0.21$ and $\sigma_{V}=0.18$ mag emphasizing the power 
of SNe\,Ia as standard candles.

\subsection{Suggested Second-Parameter Correlations for  
  SNe\,Ia in their Effects on the Determination of H$_{0}$} 
\label{sec:3_3}

     The interpretation of the supernova experiments given by 
equations~(\ref{eq:3}) and (\ref{eq:4}) has been challenged 
on the basis that there may be a range of true absolute magnitudes 
even of Branch-normal SNe\,Ia, depending on (1) details of the shape 
of the light curve (Pskovskii 1977, 1984; Phillips 1993; 
Hamuy et al.~1995; Riess, Press, \& Kirshner 1995), 
(2) intrinsic color of an individual SN (H{\"o}flich \& Khokhlov 1996), 
(3) color or Hubble type of the parent galaxy 
(Branch, Romanishin, \& Baron 1996).

     Whatever the correlations may eventually be found to be,
their total effect on $\meanl M(\max)\meanr$ is clearly small. 
The proof is that the observed dispersion in apparent magnitude 
at a given redshift in the Hubble diagram of Figure~\ref{fig:2} 
is itself so small. 

     The size of the effect suggested by Phillips (1993) was
shown to be too large by a factor of three (Tammann \& Sandage 
1995). We postulated that the problem was due to his use of 
secondary distance indicators from TF and SBF that were not 
accurate enough for the purpose. That the slope of the 
correlation of decay rate of the light curve with $M(\max)$ derived 
by Phillips is too large by a factor of between three and four 
was confirmed by Hamuy et al.~(1995, 1996).  

     The color correlation derived by Branch et al.~1996 is more
convincing, but these authors show that the effect on the Hubble 
constant, based on our six calibrating SNe\,Ia with Cepheid 
distances, is nil. They derive $H_{0} = 58 \pm 7$ from the totality of 
the data, closely the same as if no correction had been applied. 

     Whatever the final outcome will be of these suggestions that
second parameters may be needed for a better determination of 
$\meanl M(\max)\meanr$, the already tight Hubble diagrams of 
SNe\,Ia, without second parameters, show that their effect on 
$H_{0}$ will be less than 5\%. The only reason for concern 
would be a {\em systematic\/} luminosity difference between 
the seven calibrators and the 56 distant SNe\,Ia. But the former 
coming from a {\em distance\/}-limited sample and the latter from a 
{\em flux\/}-limited sample they can differ only in the sense
-- for basic principles of stellar statistics -- that the
calibrators are underluminous which causes the results in 
equation~(\ref{eq:3}) and (\ref{eq:4}) to be {\em upper limits}.

The one possibility to increase the value of $H_{0}$ is to
postulate internal absorption of the distant SNe\,Ia in spirals 
although this is against the expectations of a flux-limited sample
which is biased against all dimming effects.
In spite of this Riess, Press, \& Kirshner (1995) have proposed
non-negligible amounts of internal absorption for many SNe\,Ia
in Figure~\ref{fig:2}. However, their results raise more
problems than they solve.
(1) The already small luminosity scatter of SNe\,Ia in spirals
is closely the same as the scatter of SNe\,Ia in E/S0 galaxies;
if one ``corrects" the SNe\,Ia in spirals for internal absorption
and thereby reduces their luminosity scatter, one has to accept
the conclusion that SNe\,Ia in spirals were better standard candles
than those in early-type galaxies.
(2) The dependence of absolute magnitude on SN color $B(\max) - V(\max)$
is the same in spirals and E/S0s [$M_{\rm B} \propto 1.8\,(B-V)$ and 
$M_{\rm V} \propto 0.8\,(B-V)$ for SNe\,Ia after 1985] and is already
considerably flatter than theoretical models predict for the 
{\em intrinsic\/} correlation (H{\"o}flich \& Khokhlov 1996; 
cf. van den Bergh 1995), leaving no room for any internal absorption.
(3) The {\em total\/} absorption in $V$ proposed by Riess, Press, \&
Kirshner (1996) is less than the Galactic absorption alone
(Burstein \& Heiles 1984) by as much as $\sim\,0.24\;$mag for some 
SNe\,Ia (SN\,1992K, SN\,1993ac). For these and other reasons the
suggested absorption corrections are highly implausible.

\section{The Route based on the Distance to the Virgo cluster}
\label{sec:4}

     The second method that gives the global value of $H_{0}$ 
directly, reading a Hubble-like diagram at large redshifts       
(i.e $v \ga 10000$\kms), requires knowledge of the actual 
distance to the Virgo cluster, plus the {\em ratio\/} of distances of 
remote clusters to the Virgo cluster itself. 

     Relative distances can be found by a variety of reliable
methods, often agreed upon by both the proponents and opponents 
of both the long and short distance scales. Only the absolute 
distance of the Virgo cluster is the point of controversy. 

     Consider first the non-controversial Hubble diagram using
distances {\em relative\/} to Virgo.  
    
\subsection{The relative Hubble diagram to v  = 10000\kms}
\label{sec:4_1}

     The method was proposed and initially applied using  
distance ratios of 17 ``remote" clusters relative to Virgo 
(Sandage \& Tammann 1990; Jerjen \& Tammann 1993). The distance 
{\em ratios\/} are determined by a variety of methods 
(Tully-Fisher, D$_{\rm n}-\sigma$, first ranked cluster members). 

     The sample has now been enlarged to 31 relative distances by
adding data given by Giovanelli (1996). The complete data, listed 
elsewhere (Tammann \& Federspiel 1996), use redshifts reduced to 
the Virgocentric kinematic frame for $v < 3000$\kms\ using an 
infall velocity of 220\kms\ (Tammann \& Sandage 1985) and the 
catalog of corrections by Kraan-Korteweg (1986) of observed 
redshifts to the Virgocentric frame. The observed redshifts for    
$v > 3000$\kms\ have been reduced to the kinematic frame of the 
cosmic microwave background (CMB) using the dipole amplitude of 
630\kms. The rationale for these precepts is that there is a 
free expansion within the ``local bubble" except as decelerated by 
the Virgo complex (given by the Virgocentric corrections), and 
that the ``local bubble" is falling, in first approximation, in 
bulk toward the hot CMB pole (cf. Federspiel et al.~1994). 

\begin{figure}
\begin{center}
\leavevmode
\epsfxsize 11.5cm
\epsffile{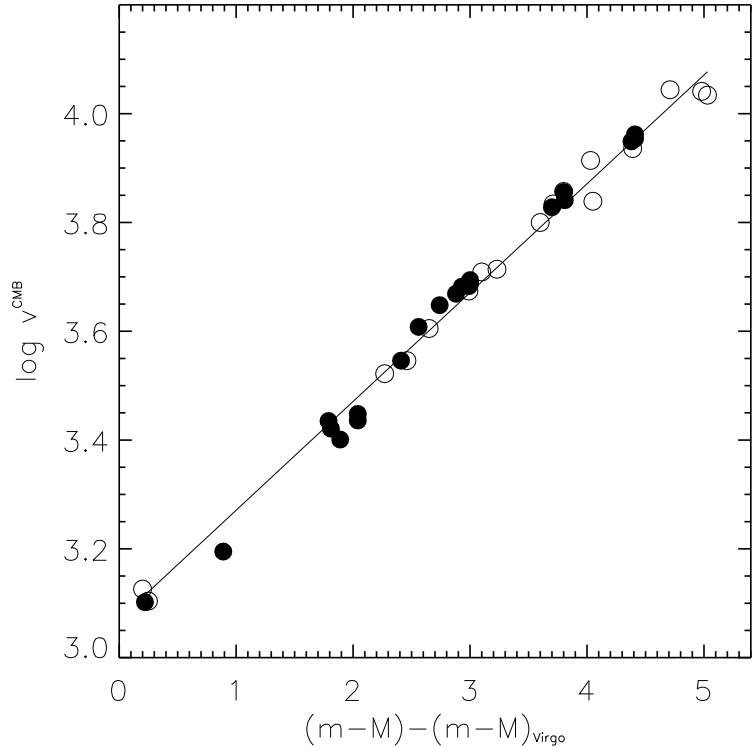}
\caption{Hubble diagram of 31 clusters with known relative 
     distances. Asterisks are data from Jerjen \& Tammann (1993). Open 
     circles are from Giovanelli (1996). Filled circles are the 
     average of data from both sources. The ordinate is log redshift 
     reduced to the CMB frame. The abscissa are the distance modulus 
     differences of each cluster relative to Virgo.}\label{fig:4}
\end{center}
\end{figure}

     The resulting relative Hubble diagram is shown in 
Figure~\ref{fig:4}. The slope of d\,mag / d\,log\,cz = 5 is forced. 
As in Figures~\ref{fig:1}\,-\,\ref{fig:3} this 
requirement for a linear expansion is well met. Furthermore, the 
small scatter 
of 0.11 mag or, read vertically, of $\Delta v/v = 0.052$ sets
an upper limit -- banning all other error sources -- 
of $\sim 260$\kms\ at a median velocity of $\sim 5000$\kms\ 
for the mean value of any (one-dimensional) random or streaming 
velocities about the ideal Hubble flow (cf. Lauer \& Postman 1994).

%

     The equation of the ridge line in Figure~4 is
\begin{equation} \label{eq:5}
  \log v({\rm CMB}) = 0.2\,[(m - M)_{\rm cl} - (m - M)_{\rm Virgo}] 
  + 3.070  \pm 0.024,  
\end{equation}
if the strongly deviating point for the Eridanus cluster (at 
delta modulus difference $\approx$ 1) is removed. 

     Note that the Virgo cluster is predicted from equation~(\ref{eq:5})
to have a cosmic velocity, freed from all random and streaming 
motions, of $1175 \pm 30$\kms. Comparing this value with the actually 
observed mean cluster velocity of $v_{0} = 922 \pm 35$ 
(Binggeli et al.~1993),
reduced to the centroid of the Local group by the precepts 
in the RSA, on obtains an ``infall" (actually retarded expansion) 
velocity of the Local Group of $253 \pm 46$\kms. We take this as 
confirmation of $v_{\rm infall} = 220$\kms\ adopted above, rather than 
the much higher values often used in the earlier literature.  

   The {\em global} value of $H_{0}$ follows from equation~(\ref{eq:5}) as
\begin{equation} \label{eq:6}
          \log H_{0} = (8.070 \pm 0.024) - 0.2\,(m - M)_{\rm Virgo}.    
\end{equation}

     For $H_0$ we need, therefore, the distance modulus of the Virgo
cluster, yet the velocity of the cluster is not needed! 
We set out in the next subsections four independent 
methods, discussed in the order of their power, that average to 
$(m - M)_{\rm Virgo} = 31.7$. Recall that advocates of the short scale    
($H_{0} \sim 80$) require $(m - M)_{\rm Virgo} = 30.9$ 
(Jacoby et al.~1992).  

     Equation~(\ref{eq:6}) can also be written in the following form
\begin{equation} \label{eq:6a}
          H_{0} = (50 \pm 3) (23.5/D_{\rm Virgo}),    
\end{equation}
where the Virgo cluster distance $D_{\rm Virgo}$ is in Mpc. This may
be compared with $H_{0} = (50 \pm 7) (21.7/D_{\rm Virgo})$ derived
earlier by tying the Virgo cluster to the relative cluster distances
then available (Tammann \& Sandage 1985).

\subsection{Virgo distance via the Tully-Fisher method}\label{sec:4_2} 

     The method using 21~cm line-widths has been applied many 
times but with variable success. Widely divergent values are in 
the literature that in some cases favor the short distance scale 
(e.g. Pierce \& Tully 1992 giving $m - M = 30.9$) and in others the 
long scale (Kraan-Korteweg et al.~1988 with $m - M = 31.6$; Fouqu{\'e} 
et al.~1990 for the same value if corrected to the modern local 
calibrators; Federspiel, Tammann, \& Sandage 1997 with $m - M = 31.7$ 
set out here).   

     It has been shown (Federspiel et al.~1994; Sandage,
Tammann, \& Federspiel 1995) that the reasons for small values of    
$(m - M)$ for Virgo (the short scale) using TF are two; (1) use of 
incorrectly small distances to the local calibrators in earlier 
papers by proponents of the short scale, and (2) neglect of the 
disastrous effect of the Teerikorpi (1987, 1990) cluster 
incompleteness bias. It can be shown that this bias produces 
errors in the modulus up to 1 mag depending on how far one has 
sampled into the cluster luminosity function {\em regardless how the 
sample is chosen\/}, if the sample remains incomplete. The modulus 
error is a strong function of the fraction of the luminosity  
function that remains unsampled (Kraan-Korteweg et al. 1988; 
Sandage et al.~1995)

     The calibration of the TF relation has been dramatically
improved by the advent of Cepheid distances with HST. There are 
now 14 Cepheid distances available for spirals suitable of the 
calibration. Detailed data with complete references to the 
extensive literature are given elsewhere (Tammann \& Federspiel 
1996; Federspiel et al.~1997) and are not repeated here. 

     A new study of the TF relation has been completed (Schr{\"o}der
1996; Tammann \& Federspiel 1996; Federspiel et al.~1997) made 
using a now complete sample of Virgo cluster spirals. Rigid 
criteria have been invoked in the selection of members over a 
tightly controlled precept as to the cluster boundaries. Many 
subtleties, not seen in earlier studies, have been found. These 
include a variation of the derived modulus on the 
wavelength of the observations (covering UBVRI), and a 
correlation of the derived modulus on the degree of hydrogen 
depletion for the spirals.   

     There is space only to quote the result, now based on the
new Cepheid calibration of the TF method, account being taken 
of the two effects just mentioned. The value adopted by Tammann \& 
Federspiel (1996) is
\begin{equation} \label{eq:7}
                  (m - M)_{\rm Virgo} = 31.66 \pm 0.15,          
\end{equation}
but they warn ``The application of the TF relation is considerably 
more intricate than often realized. It not only takes multicolor 
information for {\em complete} cluster samples, but the result is also 
sensitive to the input parameters. For example, the Virgo modulus 
in $B$ is too large by 0.07 mag (relative to the adopted mean) and 
is too short by 0.09 mag in $I$. These values may change from 
cluster to cluster depending on the color excess and the HI-
deficiency of the spiral members."
     
\subsection{Virgo cluster distance from SNe\,Ia} \label{sec:4_3}

     Galaxies associated with the Virgo cluster complex have 
produced at least eight type Ia supernovae (Sandage \& Tammann 
1995b; Tammann 1996a), including objects with older 
photometry, giving mean apparent magnitudes at maximum of $B(\max) 
= 12.10 \pm 0.13$, and $V(\max) = 12.11 \pm 0.16$. Hamuy et al.~1996 
have determined the mean apparent peak magnitudes of five blue, 
particularly well observed SNe\,Ia in Virgo as $B(\max) = 12.16 \pm 20$, 
and $V(\max) = 12.07 \pm 0.20$. Taking the larger sample, because it is 
less sensitive to depth effects in the Virgo cluster, and 
combining it with the absolute magnitude calibration via Cepheids 
observed with HST of $M_{\rm B} = -19.52 \pm 0.07$ and 
$M_{\rm V} = -19.48 \pm 0.07$ from 
equations~(\ref{eq:1}) and (\ref{eq:2}) gives 
\begin{equation} \label{eq:8}
               (m - M)_{\rm Virgo} = 31.61 \pm 0.16.       
\end{equation}

\subsection{Virgo cluster distance from globular clusters} 
\label{sec:4_4}

     The peak of the luminosity function (LF) of globular 
clusters (GC) has frequently been used as a possible standard 
candle. A new calibration of GCs in the Galaxy and in M\,31 combined 
with a compilation of published GCLFs in five Virgo ellipticals 
has led to a Virgo modulus of $(m - M) = 31.75 \pm 0.11$ (Sandage \& 
Tammann 1995a). Meanwhile, Whitmore et al.~(1995) found a very 
bright peak magnitude in $V$ and $I$ for NGC\,4486 from HST 
observations. Their data with our 1995 precepts gave $(m - M) = 
31.41 \pm 0.28$ in a critical discussion of the Whitmore et al.~
result (Sandage \& Tammann 1996). However, later data make it 
unclear that the GCs in NGC\,4486 are suitable for the experiment. 
The NGC 4486\,GC system have a bimodal color distribution in       
$V - I$, unlike any sample of coeval clusters, suggesting age 
differences and possible merger effects (Fritze-v.\,Alvensleben 
1995; Elson \& Santiago 1996). Turning a blind eye to this problem 
and averaging over all available data for Virgo cluster GCLFs gives 
$(m - M) = 31.67 \pm 0.15$. We are aware that the method may still 
face considerable uncertainties.   

\subsection{Virgo cluster distance from resolved Cepheids} 
\label{sec:4_5} 

     We now must approach the most controversial aspect of the 
disagreement between us and our critics. The first HST Cepheid 
distance of a galaxy associated with the Virgo complex was for 
NGC\,4321 (M\,100) (Freedman et al.~1994). Amidst unprecedented 
publicity with its subsequent major influence in the archive 
literature, the surprisingly small distance of 
$(m - M) = 31.2 \pm 0.2$ $(D = 17.1 \pm 1.7\;$Mpc) 
was precipitately interpreted as the distance to 
the Virgo cluster E galaxy core itself (Freedman et al.~1994; 
Mould et al.~1995; Kennicutt et al.~1995). These authors further 
adopted the large value of the cosmic velocity of the Virgo core 
of 1404\kms, unsupported by equation~(\ref{eq:5}),  to 
obtain $H_{0} = 82 \pm 17$.  

     Their precept that M\,100 itself defines the distance to the
cluster core was of course the only way the authors could 
proceed; it was the only distance they had. However, the well 
known wide spatial spread of the spirals as an envelope 
surrounding the more compact E cluster core already signaled a 
back-to-front ratio in Virgo that eventually must be, and now has 
begun to be, accounted for.  

     Proof that a large cluster depth effect exists came with the
Cepheid distance (San\-dage et al.~1996) to NGC\,4639, parent galaxy 
to the type Ia supernova 1990N. The distance modulus 
of this {\em low-velocity\/} (and hence certain cluster member) galaxy
was determined to be $(m - M)_{0} = 32.03$ $(D = 25 \pm 2.5\;$Mpc), 
which is 0.8~mag fainter than for M\,100. Two galaxies, each assumed 
to be a member of the cluster, cannot both define the distance to 
the cluster {\em core\/} when their distances are in the ratio of 1.5.  
The important distance difference between the two cluster members is 
also supported by their TF distances (Federspiel et al.~1997).

     Because spirals clearly form an extended envelope
surrounding the condensed E galaxy core, distances to many more 
spirals in the Virgo region must be determined before a ``Cepheid 
distance" to Virgo can be determined by this direct assault.\footnote{%
   It should be noted that the route through SNe\,Ia in Section~3 
   where, to be sure, the spirals NGC\,4496A, NGC\,4536, and NGC\,4639 
   are used in the SNe\,Ia calibration of 
   $\meanl M(\max)\meanr_{\rm SNeIa}$, no use is 
   made of their probable Virgo cluster connection, either as to the 
   distance to the cluster core itself, or any redshift data 
   therefrom.}   

     There is however an indirect method via a Cepheid distance
to a spiral in the Leo group where the back-to-front ratio may be 
more favorable. Tanvir et al.~(1995) have determined a Cepheid 
distance to NGC\,3368 (M\,96) and Graham et al.~(1996) have a Cepheid 
distance to NGC\,3351 (M\,95). The mean of the two is $(m - M) = 
30.22 \pm 0.12$ (increased by 0.05 mag for a zeropoint offset of HST 
photometry for bright stars, following Saha et al.~1996). 
If also the red-giant tip distance of NGC\,3379 (Sakai et al.~1996)
is considered the mean group distance becomes 
$(m - M) = 30.28$.\footnote{Cepheid observations in an other 
   Leo group member, NGC\,3627, parent of SN\,1989, by our consortium 
   are scheduled for cycle~6 of HST.}
The relative distance between the Leo group and the Virgo core is 
moderately well determined to be a modulus difference of $1.25 \pm   
0.15$ mag, based on five indicators (see Table~3 of Tammann \& 
Federspiel 1996). Adding this difference to the adopted mean 
modulus of the Leo group gives 
\begin{equation} \label{eq:9} 
           (m - M)_{\rm Virgo} = 31.53 \pm 0.21,         
\end{equation}
which we adopt in Table~\ref{tab:2} below to be the
(provisionally determined) ``Cepheid modulus to Virgo". 

\subsection{Other Methods} \label{sec:4_6}

     Other methods summarized elsewhere (Tammann  \& Sandage 1996; 
Sandage \& Tammann 1995b; Tammann 1996a) can only be 
mentioned here for lack of space. We simply list the results for 
the D$_{\rm n}-\sigma$ recalibration of Dressler's (1987) 
result by Tammann (1988) and the measurement of normal novae in 
three Virgo E galaxies (Pritchet \& van den Bergh 1987) discussed 
in the same summaries just cited.  

\subsection{Conclusion on H$_{0}$ by going through Virgo} 
\label{sec:4_7}

     The summary of the above  methods to the distance of the 
Virgo cluster is in Table~\ref{tab:2}.

%
\begin{table}
\caption{The Virgo cluster modulus from various methods} \label{tab:2}
\begin{center} \tablefontsize
\begin{tabular}{lcl}
            Method  &     {$(m - M)_{\rm Virgo}$} &   Hubble type \\
\noalign{\smallskip}
\hline 
\noalign{\smallskip}
        TF                    &  $31.69 \pm  0.15$ &     S         \\
        SNe\,Ia               &  $31.61 \pm  0.16$ &     E, S0, S  \\
        Globular Cl           &  $31.67 \pm  0.15$ &     E, S0     \\
        Cepheids              &  $31.53 \pm  0.21$ &     (E), S    \\
        D$_{\rm n} - \sigma$  &  $31.85 \pm  0.19$ &     S0, S     \\
        Novae                 &  $31.46 \pm  0.40$ &     E         \\
\noalign{\smallskip}
\hline
 \noalign{\smallskip}
                 mean         &  $31.66 \pm  0.08$ & \\
\end{tabular}
\end{center}
\end{table}
%

     The six methods give very consistent results. This is
remarkable in two respects. First, the methods include 
{\em independent\/} scales. The TF, SNe\,Ia, and Cepheid methods 
depend on the zero point of the P-L relation of Cepheids, the globular 
clusters rest on the absolute magnitude calibration of the RR 
Lyrae (controversial to be sure, but only at the 0.2 mag level; we 
have used that of Sandage 1993b), the D$_{\rm n} - \sigma$ method 
depends on the Cepheid distances to M\,31 and M\,81 but also on the 
independent size of the Galactic bulge, and the novae rely on Cohen's 
(1985) Galactic calibration. Second, the different distance 
determinations comprise both spiral, S0 and E galaxies.  

     Adopting $(m - M)_{\rm Virgo} = 31.66$  and using 
equation~(\ref{eq:6}) of Section~\ref{sec:4_1} gives
\begin{equation} \label{eq:10}
      H_{0}({\rm global}) = 55 \pm 2 \qquad\mbox{(internal error),}          
\end{equation}
for the direct method through Virgo but tied to the Machian frame 
at distances of 10000\kms\ through Figure~\ref{fig:4}. 

\section{Does going to Coma help?}\label{sec:5}

\subsection{Improper Precepts}\label{sec:5_1}

     Both the Freedman et al.~(1994) Key Project consortium and 
the Tanvir et al.~(1995) astronomers attempt to pass by the 
question of the correct Virgo or Leo cosmic redshift (i.e in the 
Machian kinematic frame). Rather they adopt an assumed 
distance modulus {\em difference} between the Coma cluster and either 
the Virgo core or the Leo group. They then determine their Coma 
distance by adding the assumed modulus differences to their Virgo 
or Leo distance and then use the observed or somehow corrected 
redshift of Coma to divide into their Coma distance to obtain 
$H_{0}$. This assumes no random motion of Coma relative to the 
cosmic frame. Whitmore et al.~(1996) follows the same procedure 
using a Virgo modulus from the globular clusters in M\,87, 
obtaining, as does Freedman et al. and Tanvir et al., high values 
of $H_{0}$ near 80.  

     It is of course a circular exercise to go through Coma if
their assumed Virgo moduli via M\,100 and the GC in NGC\,4486 are  
wrong for the reasons discussed earlier. Furthermore, one need 
not assume that Coma has no random motion relative to the cosmic 
flow; any random velocity can be determined with high 
precision from equation~(\ref{eq:5}) once any particular modulus 
difference between Coma and Virgo is assumed. The method is 
discussed elsewhere (Sandage \& Tammann 1996, Section~4). 

     For example, if the modulus difference between Coma and
Virgo is $3.72 \pm  0.09$ mag, which is the mean of the value used by 
Jerjen \& Tammann (1993) and the two values given by Dekel (1995) 
using the ``Potent" formalism for systematic motion, then the 
predicted cosmic redshift (in the Machian kinematic frame) of 
Coma from equation~(\ref{eq:5}) is 6516\kms. With 
$(m - M) = 31.66 \pm  0.08$ 
for Virgo, the Coma distance modulus is then $35.38 \pm 0.12 $ 
(or D = $119 \pm 7\;$Mpc), and the Hubble constant becomes 
\begin{equation}\label{eq:11}
              H_{0} = 6516/119 = 55 \pm 4.           
\end{equation}
The difference between the observed and cosmic redshift of 7188\kms\
and 6516\kms\ implies a one-dimensional peculiar motion of the Coma 
cluster of 672\kms with respect to the CMB.
This value is high in comparison with the mean peculiar motion of 
clusters (Section~\ref{sec:4_1}), but it compares well in size with 
the local CMB motion. 

     On the other hand, if the modulus difference was 3.80 mag
from Jerjen \& Tammann, the cosmic Coma redshift would be 6761\kms,
the distance would be $124\;$Mpc, the Hubble constant would
again be 55, and the peculiar motion would be smaller at 427\kms.
Of course the Hubble constant is identically the same by both
procedures because, by adopting the same Virgo modulus of 31.66, 
we simply move along the correlation line in Figure~\ref{fig:4}. 
This is the line for constant Hubble constant 
(Sandage \& Tammann 1996, Section~4).   

\subsection{Going to Coma Directly}\label{sec:5_2}

     A major HST experiment has been completed by Baum et al.~(1995) 
directly on the globular clusters in the off-center E 
galaxy, NGC 4481, in the Coma cluster. From the globular cluster 
luminosity function they obtain a {\em minimum\/} distance 
of $108 \pm 11\;$Mpc, or  $(m - M)_{\rm Coma} > 35.21$. 
With the observed redshift of $7188 \pm 450$\kms\ corrected for 
the {\em local\/} CMB motion (Jerjen \& Tammann 1993), then 
\begin{equation}\label{eq:12}
                       H_{0} < 67 \pm 8.            
\end{equation}

     This {\em minimum\/} distance of $108\;$Mpc is 17\% larger than 
the actual distance of 92.6$\;$Mpc set out by Whitmore et al.~(1995).  

     In this direct way through Coma, independent of Virgo, we
of course have no value for the modulus difference with Virgo 
unless we again use our assumed absolute Virgo modulus from 
Table~\ref{tab:2}. To remain independent of Virgo 
forbids to use equation~(\ref{eq:5}) to determine the cosmic 
velocity $v^{\rm CMB}$ of Coma; the peculiar motion of the cluster 
must then remain unaccounted for.

\section{Is there a route through Fornax?}  \label{sec:6}

     Despite enormous efforts over the past 30 years, even the 
distance ratio of the Fornax cluster to Virgo is only poorly 
known. A listing of the 30 investigations, given elsewhere 
(Tammann \& Federspiel 1996), separates the data by Hubble type 
into distances relative to Virgo for spirals and E/S0 galaxies. 
The difference in distance moduli varies from -0.40 mag, Fornax 
being closer than Virgo, to being more distant by + 0.70 mag. 
Many different and independent methods have been used in these 
30 investigations (for example TF, brightest cluster galaxies, 
color/luminosity correlations for early-type galaxies, the 
surface brightness/absolute magnitude relation for dE galaxies, 
globular clusters, SNe\,Ia, SBF, planetary nebulae, etc), and the 
details of the work are enormous, seen in the summary table by 
Tammann \& Federspiel (1996).   

     Setting aside questions of systematic errors between the
methods (which are largely unknown), the weighted averaged data 
give the modulus difference between Fornax and Virgo for the 
Fornax spirals as $-0.22 \pm 0.06$ mag (Fornax being closer), and 
$+0.13 \pm 0.07$ mag for the E and S0 early-type galaxies, giving a 
difference between the types of $0.35 \pm 0.09$. We are moderately 
convinced that this is not due to a difference at Virgo between 
the Virgo spirals and E/S0 types because here the two types of 
galaxies give closely the same mean Virgo distance 
(cf. Section~\ref{sec:4_7}). 

     Because of the spread in the totality of the 30
investigations, it would be premature to take the separation in 
space between the early-type and late-type Fornax galaxies to be 
real. Nevertheless it may be a warning that the Fornax complex is 
complicated, perhaps even elongated along the line of sight.    

     Velocities do not help much to define the cluster structure.
The mean of 41 E/S0/dE galaxies is $v_{220}  = 1323 \pm 48$\kms , 
reduced to the Virgocentric frame by our previous precepts 
(Tammann \& Sandage 1985). The dispersion for this sample is 
$ \sigma  = 307$\kms . The mean of 27 S/Im galaxies is 
$v_{220} = 1436 \pm 66$\kms\ with   $ \sigma  = 343$\kms. 
The statistical agreement of the mean velocities 
can be interpreted as the two galaxy types being at the same 
distance, but could also be a result of the S/Im members lying in 
the foreground and falling toward the more compact E/S0 galaxy 
cluster core.  

     The mean over all types is $v_{\rm LG} = 1366 \pm 50$\kms, 
reduced to the centroid of the Local Group by the precepts used in 
the RSA, or $v_{220} = 1338 \pm 50$\kms\ again reduced to the 
Virgocentric kinematic frame.   

     The cluster at a distance of roughly 30$\;$Mpc from Virgo 
may have a large peculiar motion of its own, 
signaled by the significantly higher mean cluster velocity 
than that of Virgo, even though they are at about the same 
distance from us. If we adopt the mean modulus difference of 0.00 mag, 
obtained by averaging all data from the 30 determinations 
mentioned earlier (there are many more determinations for E/S0 
galaxies than for spirals), equation~(5) predicts a cosmic 
velocity of 1175\kms, significantly different than the 
observed velocities quoted above.    

     The conclusion from this is that the Fornax cluster is much
less suited for the determination of $H_{0}$ than the Virgo cluster. 
The possible separation of the Fornax members by Hubble type, its 
expected non-negligible peculiar motion, and the very low weight 
of the determination of its $v^{\rm CMB}$ velocity, call for great 
caution. 

     Four exercises to derive $H_{0}$ from Fornax data illustrate
the difficulties. Here we do not carry the errors in $H_{0}$ due to 
uncertainties in various cluster velocities, but show only errors 
due to distance uncertainties.

     (1) A Cepheid distance of $18.2 \pm 1.3$ Mpc $(m - M) = 31.30 \pm
0.15$ has been announced by Silvermann et al.~(1996) for the 
exceptionally large spiral NGC\,1365 in the region of the Fornax 
cluster. If this value is confirmed by later definitive 
photometry and if it is naively taken as the distance to the 
compact E/S0 cluster core, and using $v_{220}= 1338$\kms\ as the 
correct velocity within the ``local bubble" in the Virgocentric 
kinematic frame (i.e adopting the kinematic model of the near 
expansion velocity field in Federspiel et al.~1994), one obtains 
$H_{0} = 74 \pm 6$.     

     (2) The turn-over magnitudes of the GCLFs of seven early-type
Fornax members have been compiled by Whitmore (1996).
The result of $\meanl m_{\rm to}\meanr_{\rm V}=23.80 \pm 0.08$ is 
suspicious because it implies the early-type Fornax galaxies to be 
{\em nearer\/} by $0.13 \pm 0.09$ mag than eight early-type Virgo 
galaxies which give $\meanl m_{\rm to}\meanr_{\rm V}=23.93 \pm 0.04$ 
(cf. Whitmore 1996).
If in spite of this, the Fornax value is accepted 
and combined with the absolute calibration of
$\meanl M_{\rm to}\meanr_{\rm V}= -7.62 \pm 0.20$ 
(realistic external error)
from the Galaxy and M\,31 (Sandage \& Tammann 1995a) one obtains 
$(m-M)_{\rm Fornax} = 31.42 \pm 0.22$ or $19.2 \pm 2.1\;$Mpc. 
This gives with $v_{220}=1338$\kms\ a value of $H_0 = 70 \pm 7$.
Whitmore (1996) derived $H_0 = 84$ because he calibrated 
$\meanl M_{\rm to}\meanr_{\rm V}$ on the assumption that the early-type
Virgo galaxies with known GCLFs were at the same distance as the spirals
NGC\,4321 (M\,100), NGC\,4496A and NGC\,4536, all three of which are 
particularly well resolved and now {\em known} to lie on the near side 
of the cluster, shown by the large distance of NGC\,4639.

     (3) If one combines the mean modulus difference with Virgo
of 0.00 mag with the cosmic velocity inferred from equation~(\ref{eq:5}) 
of 1175\kms, and using $(m - M)_{\rm Virgo} = 31.66$ from 
Table~\ref{tab:2}, then $H_{0}(\rm cosmic) = 55$. This, of course is the 
same for Virgo (equation~\ref{eq:10}) because the input numbers 
have been made the same by assuming the same distance.  

     (4) Given the very small dispersion in M(max) for SNe\,Ia
demonstrated earlier from Figure~\ref{fig:1}, then the best 
determination of the distance to the Fornax cluster is from the 
three SNe\,Ia produced by the early-type Fornax galaxies 
NGC\,1380 for SN\,1992A and NGC\,1316 for the two SNe\,Ia 1980N 
and 1981D. The apparent magnitudes for these prototypical 
Branch-normal SNe\,Ia are in Table~\ref{tab:3}. 
%
\begin{table}
\caption{Data for the three SNe\,Ia in Fornax cluster galaxies}
\label{tab:3}
\begin{center} \tablefontsize
\begin{tabular}{llccc}
          SN   &   galaxy  &  $B(\max)$ &  $V(\max)$ & $m - M$  \\
\noalign{\smallskip}
\hline 
\noalign{\smallskip}
        1980N  &  NGC\,1316 &  12.49  &  12.44  &  31.76 \\
        1981D  &  NGC\,1316 &  12.59  &  12.40  &  31.79 \\
        1992A  &  NGC\,1380 &  12.60  &  12.55  &  31.88 \\
\noalign{\smallskip}
\hline
\noalign{\smallskip}
               &       mean &  12.56  &  12.46  &  31.81 \\
\end{tabular}
\end{center}
\end{table}
%

     The apparent magnitudes in $B$ and $V$ are taken from the
summary by Hamuy et al.~(1996). 

     The distance moduli in the final column of Table~\ref{tab:3} are 
the  mean of the moduli in $B$ and $V$ assuming absolute magnitudes of
$M_{\rm B}(\max) = -19.32$ and $M_{\rm V}(\max) = -19.28$. 
These values have been adopted 
in the most conservative manner possible. They are 0.2 mag 
fainter than set out earlier in Section~\ref{sec:3} 
	as equations~(\ref{eq:1}) and (\ref{eq:2}). 
If second-parameter corrections are required to SNe\,Ia 
(Saha et al.~1997) 
depending on galaxy type, the data suggest that SNe\,Ia in 
early type galaxies are slightly fainter than in spirals. The 
present calibration of M(max) for SNe\,Ia (Sandage et al.~1996; 
Saha 1996) is made via Cepheids only in spirals, whereas the 
galaxies in Table~\ref{tab:3} are early types.   

Note that the exclusion of the outlying galaxy NGC\,1316 as a 
reliable cluster member has essentially no effect on the following 
conclusions, because the distance based on only SN\,1992A in 
NGC\,1380 is only slightly {\em larger\/} than the adopted mean 
distance of Table~\ref{tab:3}.

     Using $(m - M) = 31.81 \pm 0.20$  (or $D = 23.0 \pm 3\;$Mpc) 
based on Table~\ref{tab:3} with $v^{\rm CMB} = 1259$\kms\ calculated 
from equation~(\ref{eq:5}), 
a modulus difference of 0.15 mag from Virgo based on Table~\ref{tab:3}, 
and $(m - M)_{\rm Virgo} = 31.66$, gives 
\begin{equation} \label{eq:13}
                H_{0}  = 1259/23.0 = 55 \pm 6         
\end{equation}
from Fornax.

\subsection{The Route taken by Freedman et al.~(1996) through Fornax}
\label{sec:6_1}

     The route to equation~(\ref{eq:13}) is straightforward if we (a) 
adopt the precepts that SNe\,Ia have only a small intrinsic 
scatter in $M(\max)$ as proved by Figure~\ref{fig:1} and that the 
calibration is given relative to Cepheids by equations~(\ref{eq:1}) and 
(\ref{eq:2}), and (b) that the cosmic velocity of Fornax is given by 
equation~(\ref{eq:5}) using a modulus difference of 0.15 mag. Freedman 
et al. (1996) reject both of these precepts, and proceed as 
follows.    

     They assume that their Cepheid distance for the 
spiral NGC\,1365 with $(m - M) = 31.3 \pm 0.15$ is the distance to the 
Fornax cluster core. With this distance defining also the 
distance to the two parent galaxies of the three SNe\,Ia in Table 
3, they derive absolute magnitudes for these SNe\,Ia $M_{\rm B}(\max) =      
-18.74$ and $M_{\rm V}(\max) = -18.84$, contradicting the small dispersion 
relative to equations~(\ref{eq:1}) and (\ref{eq:2}) and also the direct 
evidence from the tightness of the Hubble diagram of Figure~\ref{fig:2}. 

     The result, although contradicting these external evidences,
permits them to dismiss the value of $H_{0}$ from equations~(\ref{eq:3}) 
and~(\ref{eq:4}) which were derived needing no assumptions as to pedigree 
of parenthood for the SNe\,Ia used there. 
Six of the calibrating SNe\,Ia lie in galaxies with proper Cepheid
distances. Only the seventh calibrator, SN\,1989B in NGC\,3627,
was assumed to share the distance with three other Leo group members;
its omission would have no effect on our adopted 
$\meanl M(\max)\meanr_{\rm SNeIa}$ (cf. Saha et al.~1997).

     Our response is that their precept that the distance to 
NGC\,1365, whatever its final value may be, does not define the 
distance to NGC\,1380 and NGC\,1316, just as the distance of M\,100 
does not define the distance to the Virgo cluster core, shown by 
the large Cepheid distance to the similar Virgo spiral NGC\,4639 
(Sandage et al.~1996; Saha et al.~1997).  

\section{Methods not used} \label{sec:7}

     We must mention in passing the methods based on planetary 
nebulae and surface brightness fluctuations, both of which have 
an extensive literature (Jacoby et al.~1992 for a review). We 
have not used either of these methods for several reasons. 

     Discussions elsewhere of the planetary nebulae method
(Bottinelli et al.~1991; Tammann 1993; M{\'e}ndez et al.~1993) have 
considered in a detail, not repeated here, the problem 
encountered by a sloping bright end of the luminosity function 
and its effect on the determination of distances. Whatever the 
difficulties with the method and/or its calibration, we judge the 
small modulus of the Virgo cluster at $(m - M) = 30.84$ determined 
with the method (Jacoby et al.~1990), compared with $(m - M) = 31.66$ 
from Table~2, as unrealistic.   

     Our basis of not understanding the results of the surface
brightness fluctuation method is the same. Detailed discussions 
have been set out elsewhere (cf. Tammann 1996a) giving reasons 
why the calibration of the method may need a new evaluation. In 
particular, the SBF distance to the Virgo cluster of $31.03 \pm 0.05$ 
(Tonry et al.~1997) is not only $0.63 \pm 0.09$ mag smaller than by 
all other evidence (Table 2), but it also implies a mean absolute 
magnitude of $M_{\rm B}(\max) = -18.93 \pm 0.14$ for the eight 
Branch-normal SNe\,Ia in the Virgo cluster discussed earlier. 
This is excluded by the seven Cepheid-calibrated SNe\,Ia giving 
$M_{\rm B}(\max)  = -19.52 \pm 0.07$ as in equation~(\ref{eq:1}) 
and set out in detail elsewhere (Sandage et al.~1996; Saha 1996) 
and by all existing type Ia {\em models\/} 
(see Tammann \& Federspiel 1996, their Section~4). 
The reasons remain a mystery, but the evidence is sufficient 
to make the method suspect for us.     

   
\section{Third route through field galaxies 
         corrected for selection bias}
\label{sec:8}

     The overriding power of the first two routes to $H_{0}$ 
in Sections~\ref{sec:3} and \ref{sec:4}, is that the calibrations of 
$M(\max)_{\rm SNe\,Ia}$, and the absolute distance of the Virgo cluster 
can be used to calibrate Hubble diagrams that extend far into the 
cosmic expansion field, independent of any and all local velocity 
anomalies. The traditional methods using local calibrations and 
local galaxies (for example to the limit of the RSA at $v < 3000$\kms) 
do not have that advantage. They are much more sensitive to the details 
of the local velocity field and to the effect of observational 
selection bias on flux-limited local samples in the presence of a 
much wider intrinsic dispersion of $\meanl M\meanr$ than for SNe\,Ia 
and the distance {\em ratios\/} that enter Figure~\ref{fig:4}. 

     Nevertheless, the first determinations of $H_{0}$ (Robertson
1928; Lemaitre 1927, 1931; Hubble \& Humason 1931) were made by 
calibrating $\meanl M\meanr$ for local galxies and applying that 
calibration to a general field sample, generally, to be sure, with 
no discussion of selection bias.     

     The method was improved fundamentally with the discovery by
van den Bergh (1960a,b) of a new sub-classification system based 
on ``luminosity classes" depending on the ``beauty", (or 
geometrical entropy) of galaxian images. He showed that this 
subdivision by regularity of the spiral pattern (beauty) narrowed 
the luminosity function far beyond that which would have applied 
across the entire wide morphological boxes of the original Hubble 
sequence, even within a given Hubble class.  

     Once the calibration of appropriate van den Bergh luminosity
classes could be obtained by fundamental (Cepheid) means, and/or 
by luminosity ratios established between the classes via 
relative Hubble diagrams, the local Hubble constant follows 
immediately if, but only if, the effect of observational bias can 
be determined and eliminated. 

     The problem of observational selection bias has been a major
stumbling block for every discussion of the $H_{0}$ problem using 
local galaxies. The bias has often been ignored, whereas, in fact, 
it is the reason for the difference between the short and the long 
distance scale.

     The case has been made in a series of papers devoted (a) to 
the effect of the bias, and (b) to developing methods to correct 
for it using local samples that are flux-limited rather than 
distance-limited. In every case, the corrections based either on 
what we have called ``Spaenhauer diagrams" (Sandage 1994a,b; 
Federspiel et al.~1994), or what Bottenelli et al.~(1986a,b) and  
Theureau et al.~(1997) have called the ``plateau of non-biased 
data", show that bias corrections dominate the answer.     

     Table~\ref{tab:4} summarizes the data now available on this 
way to $H_{0}$ using field galaxies, corrected by our methods for 
observational selection bias. Rather than develop here in extenso 
again the powerful properties of the Spaenhauer diagram that lead 
to the detection of selection bias and the consequent methods to 
correct for same, we simply give the references to these methods 
papers, to which should be added Federspiel et al.~(1994), 
Sandage (1995a) and Sandage, Tammann, \& Federspiel (1995). 

%
\begin{table}
\caption{$H_{0}$ from bias corrected field galaxies}
\label{tab:4}
\begin{center}\tablefontsize
\begin{tabular}{lll}
         Method       &   \multicolumn{1}{c}{$H_{0}$}      &    Source \\
\noalign{\smallskip}
\hline
\noalign{\smallskip}
Tully-Fisher, distance limited (local) &  $48 \pm  5$ & Sandage 1994b \\
Tully-Fisher, flux-limited (distance)  &    $< 60$    & Sandage 1994b \\
M\,101 look-alike diameters            &  $43 \pm 11$ & Sandage 1993c \\
M\,31 look-alike diameters             &  $45 \pm 12$ & Sandage 1993d \\
Luminosity class spirals               &  $56 \pm  5$ & Sandage 1996a \\
M\,101, M\,31 look alike luminosities &  $55 \pm  5$ & Sandage 1996aa \\
Tully-Fisher                 &  $55 \pm  5$  &  Theureau et al.~1996  \\
\end{tabular}
\end{center}
\end{table}
%

\section{Fourth route through physical methods} \label{sec:9}

     Physicists generally will not believe astronomical methods 
until they say that they understand the basis of these methods. 
On the other hand, astronomers, if they can show the viability of 
conclusions from {\em internal\/} astronomical proofs of the reality 
of particular correlations, such as (1) the P-L relation of Cepheid 
variables, (2) the existence of the main sequence in the H-R 
diagram, (3) the tight luminosity function of first ranked 
galaxies and SNe\,Ia before a deep {\em understanding\/} (in the 
physicists sense) of the correlations is at hand, will use these 
correlations (sans proof except that they work) to obtain new 
information.  

     The problem concerning $H_{0}$ is the same. Physicists,
suspicious of the somewhat intricate astronomical ladder, 
seek $H_{0}$ by purely physical methods. 

     To this end there now exist several possible ``purely"
physical methods to $H_{0}$, some of which are astounding in their 
near magic and beauty. Table~\ref{tab:5} summarizes the bulk of these 
methods, stating the results to mid 1996.    
%
\begin{table}
\caption{Distance determinations from purely physical methods}
\label{tab:5}
\begin{center}\tablefontsize
\begin{minipage}{13cm}
\begin{tabular}{llc}
       Method          &   \hspace*{1ex}$H_{0}$   &   Source\footnote{%
		Sources: (1) Bartel 1991 (2) Branch et al.~1996; H{\"o}flich \&
        Khokhlov 1996; H{\"o}flich et al.~1996; Ruiz-Lapuente 1996 (3) 
        Schmidt et al.~1994 (4) Baron et al.~1995 
		(5) McHardy et al.~1990; Birkinshaw \& Hughes 1994; Jones 1994; 
		Lasenby \& Hancock 1995 (6) Rephaeli 1995; Herbig et al.~1995 
		(7) Holzapfel et al.~1996; Lasenby 1996 (9) Turner 1996; 
		Kundi{\'c} et al.~1996 (10) Corbett et al.~1995; Nair 1995 
		(11) Lasenby 1996}
                                                             \\
\noalign{\smallskip}
\hline
\noalign{\smallskip}
Radio remnant of SN\,1979C in NGC\,4321 (Virgo)    & $54 \pm 20$ &  1 \\
Expanding photosphere and $^{56}$Ni SNe\,Ia models & $55-70$     &  2 \\
Expanding photosphere models of SNe II             & $73 \pm  6$ &  3 \\
Expanding photosphere models of SNe II             & $< 50$      &  4 \\
Sunyaev-Zeldovich effect for cluster A 2218        & $45 \pm 20$ &  5 \\
         \hspace*{3.5cm} for 6 other clusters      & $60 \pm 15$ &  6 \\
         \hspace*{3.5cm} cluster A 2163            & $68 \pm 30$ &  7 \\
         \hspace*{3.5cm} 2 clusters                & $42 \pm 10$ &  8 \\
Gravitational lenses QSO 0957 + 561                & $63 \pm 12$ &  9 \\
         \hspace*{3.28cm} B 0218 + 357             & $\sim 60$   & 10 \\
MWB fluctuation spectrum                 & $30 < H_{0} < 50(70)$ & 11 \\
\end{tabular}
\end{minipage}\end{center}
\end{table}

     Note that none of these methods support $H_{0}$ = 100. This was
the center of the argument as late as 1988 (see Paturel 1983 for 
a telling diagram).

\section{The age of the standard model cosmology} \label{sec:10}  
     
     Substantial progress has been made since 1990 in the 
question of an independent determination of ``the age of the 
universe". The cosmological test is, of course, to compare this 
``age from the big bang creation" with the inverse Hubble 
constant. 

   Here, we simply list in Table~\ref{tab:6} the various determinations
\begin{table}
\caption{Independent determinations of various ages}
\label{tab:6}
\begin{center}\tablefontsize
\begin{minipage}{12cm}
\begin{tabular}{lll}
     \multicolumn{2}{c}{Method}          &          Age(Gyr) \\
\noalign{\smallskip}
\hline
\noalign{\smallskip}
\multicolumn{3}{l}{A. Age of Globular clusters\footnote{%
   See also the contributions by M. Bolte and B. Paczynski 
   in this volume}}     \\
\noalign{\smallskip}
     \hspace*{1cm}
           &  (1) Sandage 1993bb           &  $14.1 \pm 0.3 $       \\
           &  (2) Chaboyer 1995            &  $11-21$ (total range) \\
           &  (3) Shi 1995                 &  $10-14$               \\
           &  (4) Mazzitelli et al.~1995   &  13($+2$, $-3$)        \\
           &  (5) Demarque 1996            &  $14.5 \pm 1.6$        \\
           &  (6) Weiss et al.~1996        &  $<13$                 \\
           &  (7) Caloi et al.~1996        &  $11-13$               \\
\noalign{\smallskip}
\multicolumn{3}{l}{B. Cooling time of white dwarfs 
                   in the Galactic bulge}                            \\
\noalign{\smallskip}
           &  (8) Wood 1992                &  $10-12$               \\
           &  (9) Segretain et al.~1994    &  $11.5-14$             \\
\noalign{\smallskip}
\multicolumn{3}{l}{C. Age of ``first" supernovae making the heavy}  \\  
\multicolumn{3}{l}{\hspace*{0.36cm} elements in the solar system}   \\ 
\noalign{\smallskip}
           &  (10) Cowan, Thielemann, \& Truran 1990; &             \\
           &   \qquad Thielemann 1995; &  $14.4 \pm 3$              \\
           &   \qquad    Truran 1996              &  $13.8 \pm 3$   \\
\noalign{\smallskip}
\hline
\noalign{\smallskip}
           & Adopt the ``mean minimum" values    &  13 ($+2$, $-3$) \\
           & Add gestation time of first stars   &  0.5             \\
\noalign{\smallskip}
\hline
\noalign{\smallskip}
  & {\em Minimum\/} Age of the ``Creation event" & 13.5 ($+2$, $-3$) \\
\end{tabular}
\end{minipage}\end{center}
\end{table}
of experiments that ``determine" (a) the age of the Galaxy, (b) 
the age of the chemical elements, and (c) the subsequent ``age of 
the universe", all independent of any consideration of $H_{0}^{-1}$.  

     For this audience there is no need for further comment,
except to note that with $H_{0}$ = 55, $H_{0}^{-1}$ = 18 Gyr, 
and $T_{\rm U} = 13.5$ Gyr there is no need to invoke a 
``crisis in cosmology" (sans consideration of $q_{0}$ -- 
the second of the two numbers; Section~\ref{sec:1}).

\section{Consequences} \label{sec:11}

\subsection{H$_{0}$ = 55 $\pm$ 10} 

     A summary of the four routes to $H_{0}$ discussed in 
sections~\ref{sec:3}, \ref{sec:4}, \ref{sec:8} and \ref{sec:9} 
is set out in Table~\ref{tab:7}, 
listed in what we believe is the power of each method.  

\begin{table}
\caption{Summary of the four routes to  $H_{0}$}
\label{tab:7}
\begin{center} \tablefontsize
\begin{tabular}{llc}            
  Method &  $H_{0}$    &      External error  \\
\noalign{\smallskip}
\hline       
\noalign{\smallskip}
          Cepheid calibrated     &             &              \\
          SNe\,Ia tied to        & $56 \pm 3$  & $+8$, $-10$  \\
          35000\kms              &             &              \\[1ex]
          Virgo distance tied to &             &              \\          
          10000\kms              & $55 \pm 2$  &        8     \\[1ex]
          Field galaxies         &             &              \\
          out to  3000\kms\      & $53 \pm 3$  &       10     \\
          corrected for bias     &             &              \\[1ex]
          Physical methods       & 58          &      (15)    \\
\noalign{\smallskip}
 \hline
\noalign{\smallskip}
            {\bf conclusion:}    & {\bf 55}    &   {\bf 10}     \\
\end{tabular}\end{center}
\end{table}

     The three independent routes to the global value of $H_{0}$,
namely (1) SNe\,Ia calibrated with Cepheids (a method that does 
{\em not\/} depend on the distance to the Virgo cluster) in 
Section~\ref{sec:3}, (2) the distance to Virgo with the cluster 
tied to the remote expansion field by Figure~\ref{fig:4} in 
Section~\ref{sec:4}, and (3) field galaxies corrected to 
distance-limited samples and calibrated with Cepheids, and then 
tied to the Virgocentric kinematic redshift frame in 
Section~\ref{sec:8}, give $56 \pm  3$, $55 \pm 2$, and $53 \pm 3$ 
(internal errors) respectively. Their only 
interdependence is that they rely on Cepheids (predominantly 
observed with HST), which are the least controversial distance 
indicators at present and which are reliable to better than  0.2 mag
($\pm 10$\% in distance) as discussed in Section~\ref{sec:2}. 
The three methods together make a strong case for $H_{0} = 55 \pm 10$ 
(external error). Values of $H_{0} < 40$ are equally unlikely as 
values of $H_{0} > 70$. Furthermore, the first two methods 
determine the global (Machian frame) value of $H_{0}$ directly, 
independent of all local velocity anomalies. 
None of the methods used by proponents of the short 
distance scale with $H_{0} > 70 $ have this property.    
In particular, all galaxies used by the ``Key Project''
consortium are local (Freedman et al.~1994, 1996; 
Silvermann et al.~1996; Graham et al.~1996). 
Our objections to their short distance scale
with $H_{0} \sim 70\!-\!80$ concern their precepts used to tie
their local data to the remote cosmic frame.
         
\subsection{Why is there still a controversy?} \label{sec:11b}

     We remain baffled. We see no single reason. At least six 
real points, (and a seventh as well) carry part of the burden 
against the argument by the proponents for the short distance 
scale with $H_{0} > 70$ that still dominates the literature. 
\begin{enumerate}
      \item An unwarrantly high recession velocity of the Virgo
cluster; 
     \item the unrealistic expectation to fathom the depth of the
Virgo cluster with only one resolved spiral galaxy (Freedman et 
al. 1994);
     \item the similar unrealistic expectation that the distance to
NGC 1365 (Freedman et al.~1996) in the Fornax 
cluster can recalibrate the $\meanl M(\max)\meanr$ of SNe\,Ia, when the 
geometrical aspect of the problem argues against the precept, and 
further violates the small dispersion in 
$\meanl M(\max)\meanr_{\rm SNeIa}$ proved by 
external data in the six high-pedigree cases as to parentage of 
the SNe\,Ia calibrated directly with Cepheids;   
     \item the myth of a sharp, dispersionless cutoff of the
luminosity function of planetary nebulae shells, independent of 
sample size and other factors;  
     \item reliance on the surface brightness fluctuation method
that has produced an unrealistically small Virgo modulus in 
severe conflict with Cepheid distances;  
     \item ignoring the Malmquist-like biases that always
artificially increase the value of $H_{0}$ from flux-limited samples 
of field galaxies and also from incomplete cluster samples if they 
are uncorrected for observation selection bias.  
      \item the lemming problem of follow the leader.
\end{enumerate}

\noindent {\small {\bf Acknowledgement:} 
The authors thank Dres. R. Giovanelli, M. Hamuy, B. Leibundgut, 
S. Perlmutter and A.\,G. Riess for pre-publication data.
They thank the organizers of this Confernce, 
particularly Dr. N. Turok, for
having initiated many enlightening discussions.
They are indebted to Mr. B. Reindl for having prepared
the manuscript. G.A.\,T. acknowledges the support of the 
Swiss National Science Foundation.}



\end{document}